\documentclass[aps,prb,superscriptaddress,twocolumn]{revtex4-1}
\usepackage{bm}
\usepackage{amsmath}
\usepackage{hyperref}
\usepackage{graphicx}

\begin{document}

\title{Resonant electron scattering by graphene antidot}

\author{I.V. Zagorodnev}
\email{igor.zagorodnev@gmail.com}
\affiliation{Moscow Institute of Physics and Technology, Institutskiy per. 9, Dolgoprudny, Moscow Region, 141700 Russia}
\affiliation{Kotelnikov Institute of Radio-engineering and Electronics of the Russian Academy of Sciences, 11-7 Mokhovaya St, Moscow, 125009 Russia}

\author{Zh.A. Devizorova}
\affiliation{Moscow Institute of Physics and Technology, Institutskiy per. 9, Dolgoprudny, Moscow Region, 141700 Russia}
\affiliation{Kotelnikov Institute of Radio-engineering and Electronics of the Russian Academy of Sciences, 11-7 Mokhovaya St, Moscow, 125009 Russia}

\author{V.V. Enaldiev}
\affiliation{Kotelnikov Institute of Radio-engineering and Electronics of the Russian Academy of Sciences, 11-7 Mokhovaya St, Moscow, 125009 Russia}

\date{\today}

\begin{abstract}

The edge states which were observed on a linear edge of graphene may also persist on a curved edge. We calculate the elastic transport scattering cross section on a graphene nanohole supporting the edge states. Resonant peaks in the gate voltage dependence of conductivity of graphene with such nanoholes are obtained. Position and height of the resonances are determined by the localization depth of the quasibound edge states, and width -- by their lifetime. The scattering amplitude near the resonant energies has a strong valley asymmetry. We evaluate the effect of moderate edge rippling, inhomogeneity of boundary parameter along the edge, and Coulomb effects (charged nanohole) on the edge states and show that they do not affect the presence of the resonances, but can substantially influence their position, height and width. The local density of states near the nanohole also demonstrates a resonant dependence on gate voltage.

\end{abstract}

\maketitle

\section{Introduction}

Graphene with circular nanoholes, which are often called antidots, may be used for microelectronic applications \cite{bib:Thomsen,bib:Dvorak,bib:ParkCH,bib:Power}, as a metamaterial in the terahertz range \cite{bib:Nikitin,bib:Svintsov}, or for the investigation of quantum coherence effects like the Aharonov-Bohm one \cite{bib:Shen,bib:Russo}. Even impurities and defects in graphene can be treated as antidots with a very small radius \cite{bib:Basko1,bib:Shytov,bib:Kotov}.

To investigate the electronic properties of perforated graphene structures one should first describe the edge of the sample that emerges due to perforation. Two types of graphene edges are often considered: zigzag and armchair \cite{bib:Nakada,bib:Beenakker,bib:Kotov}. In the nearest-neighbor tight-binding approximation there exists dispersionless edge states (ESs) at the ideal linear zigzag edge, and there are no ESs near the armchair one. However, it is quite challenging to control the edge orientation experimentally, and even if the edges are macroscopically smooth and oriented at some well-defined angles, they are not necessarily microscopically ordered \cite{bib:Casiraghi}. Nevertheless, the ESs were detected near the monoatomic steps on graphite surfaces \cite{bib:Niimi}, in graphene-hexagonal boron nitride interface \cite{bib:ParkJ}, and, finally, in graphene structures \cite{bib:Ritter,bib:Tao}. More comprehensive investigations predict that zigzag ESs acquire a dispersion \cite{bib:Ostaay}, depend on chemical environment \cite{bib:Fujii}, and that the ESs may exist even on the armchair edge \cite{bib:Maksimov}.

Another actual problem for the nanohole is that the edge orientation changes upon go-round the hole. To avoid the difficulty, the nanohole is sometimes replaced by a hexagon with the zigzag or armchair edges \cite{bib:Oberhuber,bib:Brun}. Similar replacement is applied to a charged defect in continuous description within the framework of the two-band Dirac model \cite{bib:Shytov}. In this model, electrons in a single valley of graphene are described by the Weyl-Dirac equation
\begin{equation} \label{eq:WeylHam}
	v \bm{\sigma} {\bf p}\psi = E\psi,
\end{equation}
where $v$ is the Fermi velocity, and $\psi = (\psi_1,\psi_2)^T$ is the two component wave function. The same equation may be used for the other graphene valley.

To describe the edge of graphene, one should supplement the equation (\ref{eq:WeylHam}) by the boundary condition (BC). Usually one uses the zigzag-like BC \cite{bib:Kotov,bib:Beenakker,bib:Shytov} $\psi_2=0$ (or $\psi_1=0$) or the Berry-Mondragon (infinite mass) BC \cite{bib:Berry,bib:Rakyta,bib:Beneventano}. We consider a general BC to describe an edge which may be relaxed, reconstructed, disordered, inhomogeneous, it also may have dangling bonds, band bending \cite{bib:Allen} or impurities \cite{bib:Wimmer}. In the Dirac model such BC was discussed in Refs. [\onlinecite{bib:McCann},\onlinecite{bib:Akhmerov},\onlinecite{bib:Basko2},\onlinecite{bib:Tkachov}] and in the single-valley approximation it has the following form \cite{bib:Ostaay,bib:VVolkov}
\begin{equation}\label{eq:BC}
	\left.\left( \psi_1+iae^{-i\varphi}\psi_2 \right)\right|_{at\; edge} = 0.
\end{equation}
Here $\varphi$ is the angle between the $x$-axes and the normal to the edge point, $a$ is a phenomenological boundary parameter characterizing the edge structure.

The BC stems from the fundamental physical requirements, namely, the absence of the normal component of the current through the edge and the time reversal symmetry, and an additional assumption of the absence of intervalley interaction (for generalization see Appendix \ref{app:IntervalleyScattering}). Therefore, it describes a general type of edge without specifying its microscopic structure. The universality is paid off by the unknown boundary parameter.

Another way to describe the edge of graphene is to add to the Weyl-Dirac equation (\ref{eq:WeylHam}) the effective potential $V(\bm{r})$, which is a combination of electrostatic and staggered potentials,
\begin{equation}
	V( \bm{r}) = \frac{m(\bm{r})}{2}[(1-a^2)\sigma_0 + (1+a^2)\sigma_z],
\end{equation}
where $m(\bm{r})=0$ in the graphene. If the ''mass'' $m(\bm{r})$ outside of graphene is much greater than the characteristic energies of electrons in the graphene (width of the conduction and valence bands), i.e. $m(\bm{r})\rightarrow\infty$, then the wave function almost does not penetrate the outside and the matching condition for the wave function on the boundary leads to the BC (\ref{eq:BC}). The sign of the parameter $a$ is determined by the sign of $m(\bm{r})$ outside of graphene. The infinite mass BC, which is the same as the BC (\ref{eq:BC}) with $a=1$, was obtained in Ref. [\onlinecite{bib:Berry}] by a similar procedure for $V(\bm{r})$ containing only the staggered potential. It should be mentioned here that the reformulation of the BC in terms of the effective potential $V(\bm{r})$ takes into account only two bands, while the other bands may affect the BC significantly.

When $|a|\neq 1$, the electron-hole symmetry is broken. This can be easily understood if one considers the heterocontact of two materials, both being described by the Dirac equation with different electron affinity and band gaps \cite{bib:BVolkov}.

The value of the parameter $a$ can be found from a comparison with experiments \cite{bib:Allen,bib:Latyshev1} or from more rigorous calculations \cite{bib:Ostaay}. It is worth noting that for any edge parameter $a$, the effective potential $V(\bm{r})$ necessarily involves the staggered potential. Bound states cannot be produced by a scalar potential, except the state with zero energy \cite{bib:Downing} or the quasistationary (or quasibound) state near the overcritical Coulomb impurity \cite{bib:Shytov}. It is the combination of the scalar and staggered potential that allows to describe the bound states.

The edges of graphene can be divided into two groups: those supporting ESs (the parameter $a$ varies smoothly and $|a| \neq 1$) and those which do not ($a$ varies greatly or $|a|=1$ in the case of constant $a$). We will consider the edges of the first type, with smoothly varying (along the edge) or constant parameter $|a| \neq 1$.

Motivated by the recent progress in fabrication and characterization of antidot nanostructures \cite{bib:Latyshev1,bib:Oberhuber}, we are aimed to show possible manifestations of the ESs, for example, in transport and scanning tunneling microscopy (STM) measurements of graphene with nanoholes. For this purpose, we calculate the transport cross-section of electrons on the nanohole in graphene and the local density of states (LDOS) near the antidot. Considerable efforts were devoted to the conductivity of graphene with scatterers described by scalar potentials \cite{bib:DasSarma,bib:Novikov,bib:Titov,bib:Heinisch,bib:Wu}. A real (impenetrable for carriers) hole cannot be described by a scalar potential, because electrons freely penetrate inside the scalar potential region. As mentioned above, the BC (\ref{eq:BC}) necessarily contains the effective staggered potential. Scattering by staggered potentials only was considered, for example, in Refs. [\onlinecite{bib:Thomsen},\onlinecite{bib:Power},\onlinecite{bib:Masir}]. To the best of our knowledge, the combination of scalar and staggered potential has not been analyzed yet. We will show that it results in the existence of the quasistationary ESs localized near the nanohole. In turn, it leads to the peculiar peaks in the dependence of transport cross section and the LDOS on the electron energy and, therefore, in conductivity and tunnel current vs. gate voltage. The resonances in the transport cross-section correspond to the elastic scattering of electron by the quasistationary ESs. We will analyze the peaks and their robustness against inhomogeneity of the parameter $a$ and Coulomb effects.

The paper is organized as follows. Firstly, we recall the spectrum of the ESs on a graphene half-plane (Sec. II) and discuss its robustness. Then, we analyze the quasistationary states on the graphene antidot (Sec. III). In Sec. IV and V, we calculate the transport cross section and the LDOS respectively, which are our main results, represented in Figs. \ref{Fig:CS+polar} and \ref{Fig:LDOS}. In Sec. VI, we consider the scattering by a charged antidot. Conclusions are made in the final section VII.

\section{Edge States on graphene half-plane}

First, we recall the solutions of the Weyl-Dirac equation (\ref{eq:WeylHam}) on the graphene half-plane $x \geq 0$ supplemented by the BC (\ref{eq:BC}) with a constant parameter $a$ \cite{bib:VVolkov,bib:Ostaay}. Momentum  along the boundary $\hbar k_y$ measured from the projections of the valley centers on the edge ($\pm \hbar K_{0y}$) is a good quantum number. The bulk wave function is a sum of incident and reflected plane waves. The bulk spectrum is located in the energy domain $|E|\geq v\hbar|k_y|$, while the ES spectrum is outside this region. The ES band in the $E(k_y)$-plane represents the rays starting from $\pm \hbar K_{0y}$ and described by the dispersion equations
    \begin{equation} \label{eq:HalfPlaneSpectr}
        E_s = s\hbar v \frac{2a}{1+a^2} k_y, \quad sk_y(1-a^2) > 0.
    \end{equation}
$s = \pm 1$ is the valley index. The ES wave function exponentially decay away from the edge $\psi (x) \sim e^{-sx(1-a^2)k_y/(1+a^2)}$. 

It is important that the ES band is ''chiral'' (not symmetric about the center of a valley), but it is symmetric about the center of the edge Brillouin zone in agreement with the time reversal symmetry of spinless system. This fact will lead to important consequences for scattering by the antidot.

Inclusion of the intervalley interaction at the edge leads to qualitatively the same results, see Appendix \ref{app:IntervalleyScattering}. Besides, for a wide class of the translationally invariant edges, for example zigzag one, the distance between the projections of the valley centers $2|K_{0y}|$ is large in comparison with the electron momentum, hence, the edge intervalley interaction is negligible \cite{bib:Ostaay}. For large electron energy, the edge intervalley interaction is important and results in connection between the rays of the ESs from the different valleys, see Fig. \ref{Fig:ESspectra}b.

Moderate bending of graphene sheet near the edge, which probably took place in Ref. [\onlinecite{bib:Tao}], can be included by means of renormalization of the parameter $a$, see Appendix \ref{app:rippling}.

The feature of the graphene band structure is that the ESs always coexist with the bulk states. It means that they must be quasistationary with a finite lifetime, due to the probability of decay into the bulk. The edge roughness is one of the possible reasons of the decay. For small wave vectors, we evaluate this probability using the Fermi's Golden Rule as (Appendix \ref{app:roughness})
\begin{equation}\label{eq:DecayProbability}
	w \propto k_y^4,
\end{equation}
which implies that for small wave vectors the ESs near a linear edge are quasistationary. Besides, it is expected that any deviation from linearity of the edge results in finite life time of the ESs. In the next section we will directly show it for a circular hole.

\section{Quasistationary edge state at graphene nanohole}

\begin{figure} 
	\includegraphics[width=8cm]{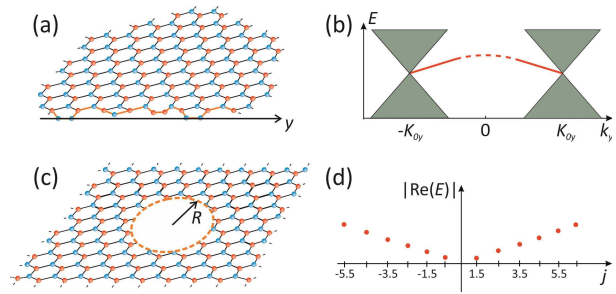}
	\caption{\label{Fig:ESspectra} (a) Sketch of the graphene half-plane (b) Energy spectrum $E(k_y)$ of the graphene half-plane. Here $k_y$ is the momentum along the edge measured from the center of the edge Brillouin zone. The shaded region corresponds to the bulk states, the rays starting in the valley centers -- to the edge states. In the absence of intervalley interaction, the rays are infinite while it inclusion results in the ''interaction'' of the rays near the Brilluoin zone center shown by the dashed line. (c) Sketch of the graphene antidot (d) Energy spectrum of the quasistationary edge states localized near the antidot (only the real part of energy is shown). The sign of angular momentum $j$ is coupled to the valley index.}
\end{figure}

We consider the ESs at a circular hole, depicted schematically in Fig. \ref{Fig:ESspectra}c. The total angular momentum $j = l+1/2 = \pm 1/2, \pm 3/2 ...$ is conserved (we will call $l$ orbital angular momentum), therefore, in the polar coordinate system ($r, \varphi$) the wave function $\psi\propto \exp(ij\varphi)$. From here, without loss of generality, we will consider negative electron energy $E<0$. Introducing the wave vector $k=-E/(\hbar v)$, one can obtain the Bessel equation for each component of the wave function. For example, the equation for the first component is
\begin{equation}
	r^2\psi_1 ''(r)+ r \psi_1 '(r) + \left[(kr)^2-l^2\right]\psi_1 = 0.
\end{equation}
The solutions corresponding to the bulk states are merely the Bessel functions or their combinations. There are no localized stationary states. However, the quasistationary states can exist. To show this, we consider complex energy $E = E'+iE''$ assuming $E'' \ll E'$, and $E''>0$. The imaginary part of the energy $E''$ determines the lifetime of the quasistationary state with energy $E'$. Since the wave function of the state is an outgoing wave, we choose it in the following form
\begin{equation}\label{eq:QuasiStationaryWF}
	\left( \begin{array}{cc}
		\psi_1(r,\varphi) \\
		\psi_2(r,\varphi)
	\end{array} \right) 
	= 
	C\left( \begin{array}{cc}
	e^{il\varphi}H_l^{(2)}(kr) \\
	-ie^{i(l+1)\varphi}H_{l+1}^{(2)}(kr)
	\end{array} \right),
\end{equation}
where $H_l^{(2)}(x)$ is the Hankel function of the second kind, and $C$ is the normalization constant. For positive energy $E'>0$, the Hankel function of the first kind must be used. Substituting the wave function (\ref{eq:QuasiStationaryWF}) into the BC (\ref{eq:BC}), we obtain the dispersion equation
\begin{equation}\label{eq:ADdispersion}
	H_l^{(2)}(kR) = -aH_{l+1}^{(2)}(kR).
\end{equation}

Expanding the Hankel functions into the Laurent series about $kR = 0$, for $l\neq 0$ we find the low energy spectrum ($|la| \ll 1/2$) of the qusistationary ESs:
\begin{equation} \label{eq:SpectrAD}
	kR \approx -2sla + i \frac{2\pi |a| (|la|)^{2l}}{\left[(l-1)!\right]^2}, \quad sla < 0.
\end{equation}
The equation describes the electron spectrum for both valleys. The spectrum of the ESs in each valley is equidistant. We stress that the ESs are chiral again, as the sign of the ES total angular momentum $j$ depends on the valley index $s$, see second part of the equation (\ref{eq:SpectrAD}). It is positive in one valley, and negative in the other one. For small parameter $a$, the real part of the energy can be obtained quasiclassically from (\ref{eq:HalfPlaneSpectr}) by replacing the momentum $k_y$ with $l/R$. 

The energy of the ''ground'' ES with $l=0$ is determined by the transcendental equation
\begin{equation} \label{eq:SpectrADl=0}
	|a| = -kR\ln\frac{kR}{2}.
\end{equation} 
However, the ground state has a long lifetime only for $kR \ll 0.1$, otherwise, $E''$ approximately equals $E'$. We will directly show it in the next section. The spectrum of the ESs including both graphene valleys is shown in Fig. \ref{Fig:ESspectra}d.

The ESs can be detected in measurements of conductivity. One of the ways to take into account the contribution of the ESs is to calculate the transport cross section and the LDOS, which will be our foregoing goal.

\begin{figure}
	\includegraphics[width=8cm]{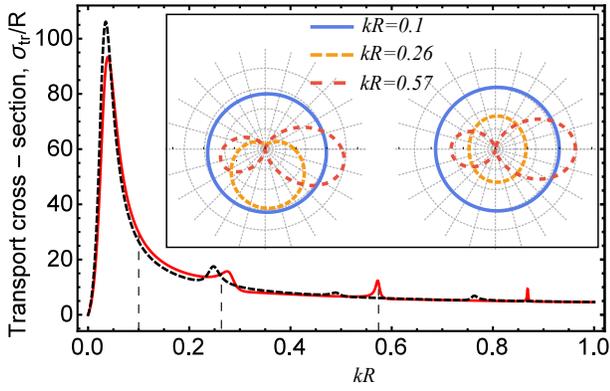}
	\caption{\label{Fig:CS+polar} Dependence of the transport cross section $\sigma_{tr}$ on the electron wave vector $k=|E|/\hbar v$ for scattering by the nanohole with raidus $R$ and the constant boundary parameter $a=-0.15$ (red solid line) and $a(\varphi) = -0.15 \pm 0.04\cos\varphi$ (black dashed line). The resonances emerge due to the scattering by the qusistationary edge states. Inset: (left) Polar plot of the scattering amplitude in a valley for $a=-0.15$ and different energies. The energies are highlighted on the main figure by the vertical dashed lines. An electron is incident on the antidot from the left. (right) Total scattering amplitude, taking into account the contribution of two valleys.}
\end{figure}

\begin{figure}
	\includegraphics[width=8cm]{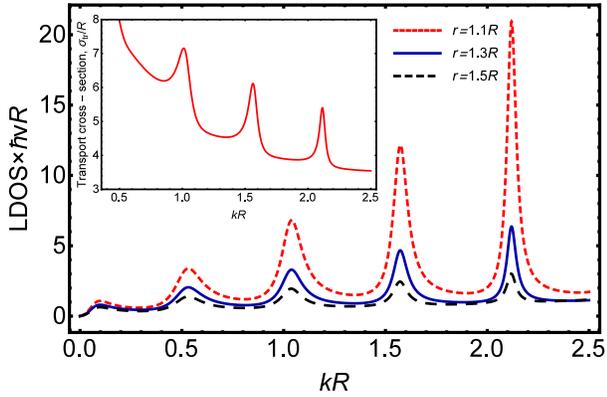}
	\caption{\label{Fig:LDOS} Dependence of the LDOS at different distances $r$ from the center of the circular hole with radius $R$ on electron energy ($k=|E|/\hbar v$) for the constant boundary parameter $a=-0.3$. Resonant peaks are clearly visible on the free LDOS trend $2k/(\pi \hbar v)$. Inset: The transport cross section $\sigma_{tr}$ vs. electron energy for the same value of the parameter $a$.}
\end{figure}

\section{Scattering by neutral antidot}
We first consider the scattering of a plane wave by the nanohole. We represent the wave function as a combination of the normalized plane wave propagating in the $x$ direction and the scattered cylindrical waves
\begin{equation} \label{eq:ScaterWF}
	\psi_{scat} =\left( \begin{array}{cc}
        1 \\
        1
    \end{array} \right) \frac{e^{-ikx}}{\sqrt{2}} + 
	\sum\limits_{l=-\infty}^{\infty} C_l \frac{e^{il\varphi}}{\sqrt{2}} 
	\left( \begin{array}{cc}
        H_l^{(2)}(kr) \\
        -ie^{i\varphi}H_{l+1}^{(2)}(kr)
    \end{array} \right).
\end{equation}
In the case of $E>0$ one must choose the $H_l^{(1)}(x)$ Hankel function.

The behaviour of the scattered wave function at large distances determines the scattering amplitude
\begin{equation}
\label{eq:scattampl}
	f(\varphi) = \sqrt{\frac{2}{\pi k}} \sum\limits_{l=-\infty}^\infty C_l \exp\left( il\varphi+i\frac{l\pi}{2} + i\frac{\pi}{4} \right).
\end{equation}

Expanding the incident plane wave into the series of cylindrical waves and substituting the full wave function into the BC (\ref{eq:BC}), we obtain
\begin{equation}
\label{eq:Cl}
	C_l = -(-i)^l\frac{ J_l(kR)+aJ_{l+1}(kR) }{ H_l^{(2)}(kR)+aH_{l+1}^{(2)}(kR) },
\end{equation}
where $J_l(x)$ is the Bessel function of the first kind.

It follows from (\ref {eq:Cl}) that $C_{l+1} \neq C_{-l}$. Therefore, introducing the scattering phase $e^{2i\delta_j}=1+2i^lC_l$ ($j=l+1/2$), one can show that $\delta_j \neq \delta_{-j}$. This results in the asymmetry of the scattering amplitude with respect to the replacement $\varphi \rightarrow -\varphi$, i.e $f(\varphi) \neq f(-\varphi)$ in one valley. The reason of this asymmetry is the lack of the time reversal symmetry in one valley, which was mentioned in Secs. II and III. The ES in a given valley has positive $j$ and rotates counterclockwise, in the other one it has negative $j$ and rotates clockwise. This chirality of the ES with respect to the valley index leads to the asymmetry of the scattering amplitude. It somehow resembles the classical Magnus effect and this skew-symmetric scattering must lead, in turn, to the valley Hall effect. Moreover, this valley Hall effect should be resonant, because the magnitude of the asymmetry strongly depends on the energy and has maximum value in the vicinity of the ES energy. At the same time, the total scattering amplitude including the contributions from both valleys is symmetric, because the time reversal symmetry is restored, see inset in Fig. \ref{Fig:CS+polar}.

It is also important to note, that the backscattering is suppressed for the scattering by scalar potentials. However, the antidot is described by the combination of scalar and staggered potentials thus allowing the backscattering.

To show a possible manifestation of the ESs in the resistivity, we calculate the transport cross section on the antidot
\begin{equation}
	\sigma_{tr} = \int\limits_0^{2\pi} \left( 1 - \cos\varphi \right) \left|f(\varphi)\right|^2d\varphi = \frac{4}{k}\sum\limits_{l=-\infty}^\infty\left( |C_l|^2 - {\rm Im}(C_lC_{l+1}^*) \right).
\end{equation}
The energy dependence of the transport cross section is shown in Fig. \ref{Fig:CS+polar} and inset in Fig. \ref{Fig:LDOS}. The resonant peaks in these figures emerge because of the resonant electron scattering by the quasistationary ESs and, therofore, are almost equidistant in accordance with the ESs spectrum obtained in the previous section.

Let us start analysis of the energy dependence of the cross section with the low energy scattering $kR \ll 2|a|$ and for small parameter $|a| \ll 1$. In this case, the scattering is symmetrical (s-scattering) and it is the ground ES that mainly contributes to the scattering cross section. Expanding the Bessel functions, we find
\begin{equation}
	C_0 \approx -(-i)^l\frac{1}{ 1+i\frac{2}{\pi}\left[ \frac{a}{kR} - \gamma - \ln\left( \frac{kR}{2} \right) \right] },
\end{equation}
where $\gamma$ is the Euler-Mascheroni constant. The cross section has the form
\begin{equation}\label{eq:CSl0}
	\sigma_{tr} \approx \frac{4\pi^2 kR^2}{ \pi^2k^2R^2+4\left[kR \ln\left(\frac{kR}{2}\right)-a\right]^2 }.
\end{equation}
Formula (\ref{eq:CSl0}) resembles equation (21) of Ref. [\onlinecite{bib:Basko1}] for a general low energy scattering with the ''scattering length'' $R/a$.
 
The cross section (\ref{eq:CSl0}) tends to zero for small energies if $a \neq 0$. To clarify this, one can use the scattering phase $\delta_{1/2}$, which is proportional to $k$ for small energies as in a general theory of non-relativistic scattering \cite{bib:LL}. Then the elastic cross section $\sigma_{el} \propto |1-e^{2i\delta_{1/2}}|^2/k \propto k$, which implies that inelastic scattering dominates for small $k$. The inelastic scattering phases are complex, and one should expect that $\delta_{1/2}$ still will be proportional to $k$ with some complex coefficient \cite{bib:LL}. The total cross section which includes elastic and inelastic channels will be finite, $\sigma_{tot} \propto [1-{\rm Re}(e^{2i\delta_{1/2}})]/k \approx {\rm const}$.

The energy of the first cross section maximum approximately corresponds to the ground ES energy (\ref{eq:SpectrADl=0}), the height of the peak is  $\sigma_{tr} = 4/k_0$, the full width at half maximum $\Delta = \pi a/[\pi^2+4\ln^2(k_0/2)]$, where $k_0$ is the root of the equation (\ref{eq:SpectrADl=0}). It follows from these equations that the ground state with $l=0$ has long lifetime only if $kR \ll 2\exp(-\pi)\approx 0.1$.

To analyze other resonances in the transport cross-section we consider the energies in the vicinity of the qusistationary ES with an orbital angular momentum $l_0$, i.e. $kR \approx 2|al_0| \ll 1$. Then, it is the coefficient $C_{l_0}$ that has a resonant energy dependence 
\begin{equation} 
	C_{l_0} \approx -\frac{1}{1-\frac{il_0!(l_0-1)!}{2\pi} \left( \frac{2}{kR} \right)^{2l_0+1} \left( kR + 2al_0 \right)} \approx -1,
\end{equation}
while other $C_l$'s behave smoothly. The height of the peak (measured from the background determined by other $C_l$'s) approximately equals $2R/|al_0|$. Thus, for sufficiently small $|al_0|$, it can be several times larger than $2R$, the geometrical cross section of the hole. The peak has the Lorentzian shape, while its width is determined by the lifetime of the ES.

Let us now estimate the contribution of the scattering by antidots to the net resistivity of a sample, using the parameters from experiments, Refs. [\onlinecite{bib:Latyshev1},\onlinecite{bib:Latyshev2}]: the Fermi energy $E_F \approx 20$ meV$\gg kT$, the concentration of the antidots $N \approx 10^{10}$ cm$^{-2}$. We use the Drude formula (if the Fermi energy is sufficiently far away from the Dirac point) with the scattering time $\tau =  1/(N\sigma_{tr} v)$ and the transport cross section $\sigma_{tr} \approx 2R \approx 20$ nm and obtain the 2D resistivity
\begin{equation}
	\rho \approx 2RN\cdot\frac{\pi v\hbar^2}{E_Fe^2} \sim 10^2 \; {\rm Ohm},
\end{equation}
which is a measurable value. Since the height of the transport cross section resonances can greatly exceed $2R$, the resonant resistivity can be much higher than $100$ Ohm and the ESs can be detected. Such measurements allow to estimate the characteristics of the ESs and the value of the boundary parameter.

Thus far we considered constant boundary parameter $a$, though it may depend on the orientation of the edge which varies upon go-round the circular nanohole. Therefore, now we simulate this situation consider inhomogeneity of the parameter $a$ along the edge, i.e. $a=a(\varphi)$. Expanding it in the Fourier series $a(\varphi) = \sum a_ne^{in\varphi}$ and substituting the wave function (\ref{eq:ScaterWF}) into the BC (\ref{eq:BC}), we find
\begin{multline} \label{eq:NonhomoA}
	\sum\limits_n a_n\left[ (-i)^{l-n}J_{l-n+1}(kR) + H_{l-n+1}^{(2)}(kR)C_{l-n} \right] + \nonumber \\ + H_l^{(2)}(kR)C_l + (-i)^lJ_l(kR) = 0, \quad \forall l.
\end{multline}
It is merely an infinite matrix equation on the coefficients $C_l$. In the simplest case of harmonically modulated parameter $a(\varphi) = a_0 + 2a_1\cos\varphi$ ($|a_1|\ll|a_0|$) one can cut off the resulting tridiagonal matrix for some sufficiently large $l$ and solve the reduced system. Numerical results for $a_0=-0.15$ and $a_1=0.04$, presented in Fig. \ref{Fig:CS+polar}, demonstrate a significant change in the position and magnitude of the resonances away from the Dirac point and, at the same time, the robustness of the resonance effect against moderate inhomogeneity of the boundary parameter $a$.

\section{Local density of states near nanohole}

We have shown that the antidot ESs manifest themselves in the energy dependence of the transport cross-section, and, therefore can be detected by measuring the conductivity of graphene sample with such antidots. However, high identity of the antidots is required for such an experiment and other types of scatterers could mask the effect. Another possible way to detect the ESs free of these drawbacks is to measure the LDOS near a hole. To this end, we calculate the LDOS near the nanohole. In the following calculations, we impose non restrictions on the partial (with total angular momentum $j$) wave function at the infinity
\begin{equation}
	\psi_{k,l} = A_l(k)e^{il\varphi} \left( \begin{array}{cc}
	J_l(kr)+B_l(k)Y_l(kr) \\
	-ie^{i\varphi}\left[ J_{l+1}(kr)+B_l(k)Y_{l+1}(kr) \right]
	\end{array} \right),
\end{equation}
where $Y_l(x)$ is the Bessel function of the second kind, $A_l(k)$ is the normalization coefficient, and $B_l(k)$ is obtained by substituting the wave function into the BC (\ref{eq:BC}):
\begin{equation}
	B_l(k)=-\frac{aJ_{l+1}(kR)+J_l(kR)}{Y_l(kR)+aY_{l+1}(kR)}.
\end{equation}

To find  $A_l(k)$, we use the normalization rule for the eigenfunctions of a continuous spectrum $\int\limits_R^\infty\psi_{k',j'}^+(r,\varphi)\psi_{k,j}(r,\varphi)d^2r=g\delta(k-k')\delta_{jj'}$, where $g$ is dimensional parameter determined from the free (without a hole) DOS per unit area. Then, the LDOS including spin and valley degeneracy is
\begin{multline}
\label{eq:LDOS}
	\rho (k,r)=\frac{k}{\pi\hbar v}\sum\limits_{l=-\infty}^\infty \frac{1}{1+\left|B_l(k)\right|^2}\left[\left|J_l(kr)+B_l(k)Y_l(kr)\right|^2\right.\\
	+ \left. \left| J_{l+1}(kr)+B_l(k)Y_{l+1}(kr) \right|^2	\right].
\end{multline}
The energy dependence of the LDOS is shown in Fig. \ref{Fig:LDOS} for $a=-0.3$. Again, it demonstrates the resonances for the energies in the vicinity of the ES energies. If $kR$ tends to zero, then all coefficients $C_l \rightarrow 0$, and we obtain the free DOS per unit area, $\rho_{free}(k) = 2k/(\pi \hbar v)$. 

The spatial dependence of the LDOS for any energy demonstrates a power low decrease at small distances and the Friedel oscillations far away from the nanohole without any resonant peaks, however, the magnitude of the LDOS near the nanohole significantly (resonantly) depends on energy. For this reason, we analyze below only the energy dependence of the LDOS near the nanohole.

For energies below the ground state energy, $k<k_0$, for $a \ll 1$ and $r \sim R$, the main contribution to the LDOS comes from the terms with $l=0,-1$ ($j=\pm 1/2$)
\begin{equation}
\label{eq:LDOSsmallk}
\rho (k,r) \simeq \rho_{free}(k)\left(1+\frac{R^2}{2a^2r^2}\right).
\end{equation}
It is considerably greater then the free DOS, increases with decreasing of the parameter $a$ and diverges when $a\rightarrow 0$, which is the typical for the zigzag case. For the linear ''reczag'' edge, the LDOS demonstrates qualitatively the same behaviour \cite{bib:Ostaay}.

If the energy is close to the ground state energy $k \simeq k_0$, then the height of corresponding peak is
\begin{equation}
\label{eq:LDOS0}
\rho (k_0,r) \simeq \frac{4}{\pi^3 \hbar v k_0r^2}.
\end{equation}
In the vicinity of the ES energy with $l_0$, $kR \simeq 2|a|l_0 \ll 1$, the coefficient $B_{l_0}$ is resonantly large, while the other $B_l$'s are small. Then, the height of the $l_0$th peak in the LDOS is
\begin{equation}
\label{eq:LDOSl0}
\rho \simeq \frac{2}{\pi^3\hbar v R}\frac{{l_0!}^2}{(|a|l_0)^{2l_0+1}}\left(\frac{R}{r}\right)^{2l_0+2}.
\end{equation}

According to Eqs. (\ref{eq:LDOS0}, \ref{eq:LDOSl0}) the height of all resonances decreases while moving away from the nanohole with a power law profile. The greater the resonances number, $l_0$'s, the stronger is the spatial decay of the LDOS. For a fixed distance from the antidot each subsequent energy peak is higher than the previous one. This behavior of resonances in the LDOS obtained in low energy limit $kR \ll 1$ is in a qualitative agreement with the results for wider energy range represented in Fig. \ref{Fig:LDOS}.

\section{Scattering by charged antidot}

In this section we consider scattering by a charged antidot. This charge could appear naturally, for example, in graphene sample deposited onto a substrate which possesses a large number of charged impurities \cite{bib:Chen}, or artificially by deposition of dimers \cite{bib:Wang}. The antidot may also become charged while electrons occupy the ESs. Our aim is to calculate the renormalization of the ESs energy spectrum and modification of the transport cross section due to the Coulomb effects.

The presence of the charge $Q$ in the center of the antidot leads to the presence of an extra Coulomb term $-eQ/r$ in the Weyl-Dirac equation (\ref{eq:WeylHam}). The edge of the antidot is still described by the BC (\ref{eq:BC}) with the constant parameter $a$.

The spectrum of quasistationary ESs is determined by the vanishing of incoming wave. Introducing dimensionless charge $q=eQ/\hbar v$, in the low-energy limit ($kR \ll 1$) and under conditions $|q| \ll 1$,  $ la<0$ for $l\neq 0$ we obtain the spectrum
\begin{multline} \label{Eq:ESSwithQ}
kR \simeq -2al+\frac{l}{l+1/2}q +\\+i\frac{|\Gamma(l+1/2-iq)|^2}{8l\Gamma^2(2l)}e^{-\pi q} \left(-4al+\frac{2l}{l+1/2}q\right)^{2l+1}.
\end{multline}

For $l=0$, the real part of $k$ is determined by
\begin{equation} \label{Eq:ESSwithQl=0}
kR \simeq k_0R-\frac{q}{1+a/k_0R},
\end{equation}
where $k_0$ is the solution of the equation (\ref{eq:SpectrADl=0}). The details of the calculations are presented in Appendix \ref{app:coulomb}.

According to Eqs. (\ref{Eq:ESSwithQ}, \ref{Eq:ESSwithQl=0}), the wave vector $k$ of quasistationary ESs decreases (increases) for negatively (positively) charged antidot when the absolute value of charge increases. At the same time, the energy of quasistationary ESs increases (decreases), because $E=-\hbar v k$. This result is intuitively clear, because negatively (positively) charged antidot repels (attracts) electons, and, therefore, the energy of the stationary states increases (decreases).

\begin{figure}[ht]
	\includegraphics[width=8cm]{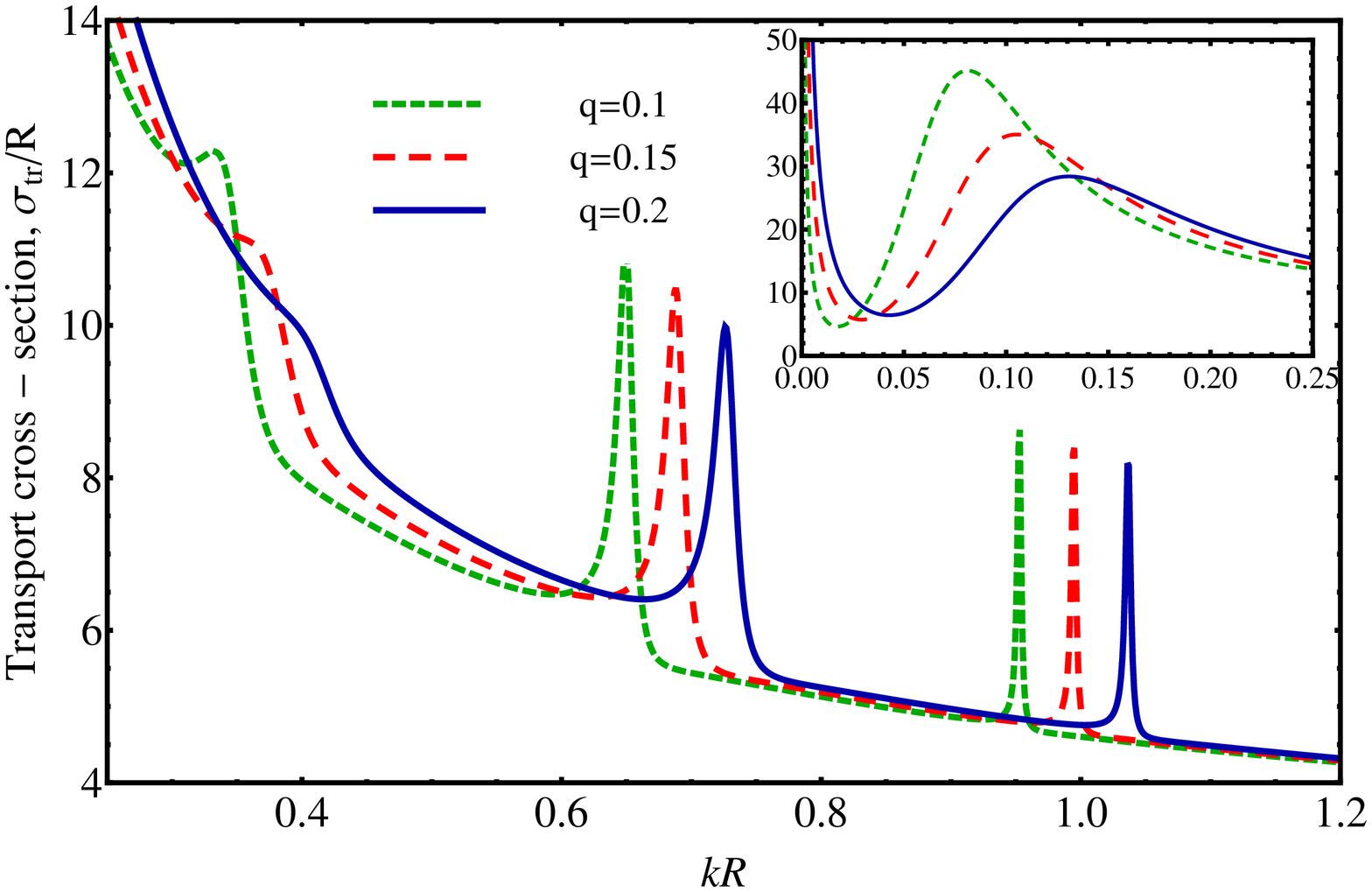}
    \includegraphics[width=8cm]{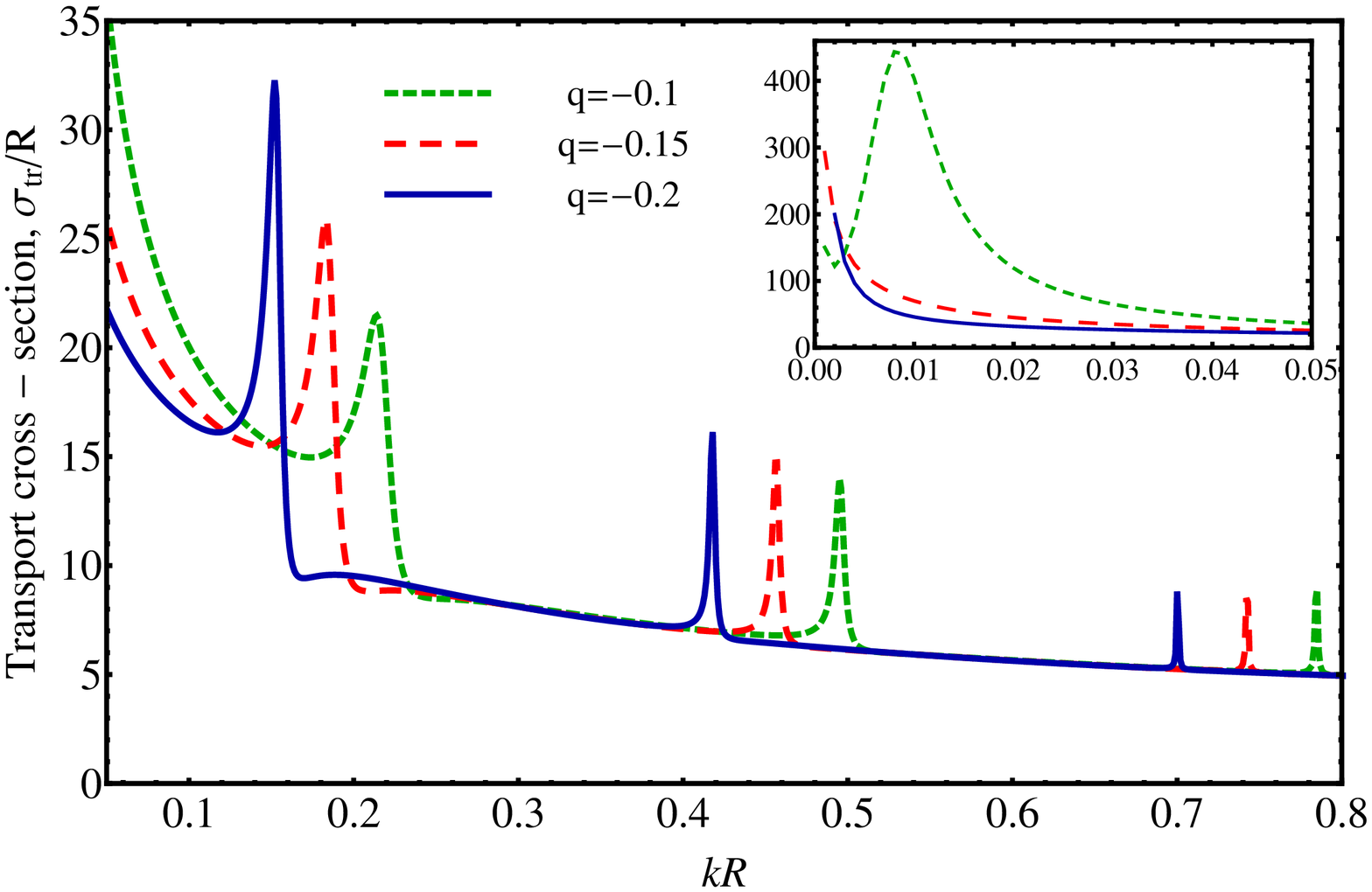}
	\caption{\label{Fig:CSwithQ} Dependence of the transport cross section $\sigma_{tr}$ on energy for the scattering by positively (upper panel, $q>0$) and negatively (lower panel, $q<0$) charged antidot and $a=-0.15$. When $|q|$ increases, the resonant peaks, which position is determined by the ESs spectrum, shift to the right for positively charged antidot, and to the left if charge is negative in agreement with Eqs. (\ref{Eq:ESSwithQ},\ref{Eq:ESSwithQl=0}). The width of the peaks depends on the ESs lifetime and decreases when the antidot is negatively charged.} 
\end{figure}

Similarly to the case of uncharged antidot, peculiar peaks corresponding to the resonant scattering by quasistationary ESs appear in the energy dependence of the transport cross section, Fig. \ref{Fig:CSwithQ}. For negatively (positively) charged antidot these peaks  shift to the left (right), when the absolute value of charge increases, in agreement with the renormalization of ESs spectrum discussed above. The width of the peaks determined by the imaginary part of the energy is also in qualitative agreement with (\ref{Eq:ESSwithQ}): at negative $q$ the imaginary part of the energy decreases when $|q|$ increases and the peaks become narrower, which corresponds to more stationary states; at positive $q$, the imaginary part of the energy increases when $|q|$ increases, and the peaks become wider.

\section{Conclusion}

We have described the electronic properties of graphene with impenetrable (for carriers) nanoholes supporting the localized (edge) states. The energies of these states are almost equidistant. We have demonstrated a strong asymmetry of the scattering amplitude near the resonances in a given valley. It must lead to the resonant valley Hall effect. The resonant scattering of graphene electrons occurs when its energy coincides with the edge state energy. Moderate edge rippling, inhomogeneity of the boundary parameter and charge of the nanohole generally do not influence the very presence of the effect, but can substantially influence the position of the resonant energies as well as the width and height of the resonances.

One way to detect these resonanses is to measure the conductivity of graphene sample with an array of identical nanoholes, another one is to measure the LDOS near an antidot. We have calculated the LDOS and shown that its dependence  on the electron energy near the hole also demonstrates resonances emerging due to the edge states.

The intervalley interaction at the curved edge of the nanohole and strong inhomogeneity of the edge parameter $a$ (or parameters if edge intervalley interaction is included) are the subject of the further studies

\section*{Acknowledgment} We are grateful to Prof. V.A. Volkov for the fruitful discussions. This work was supported by the Russian Foundation for Basic Research (projects 14-02-31592 and 14-02-01166), and the Dynasty Foundation.

\appendix

\section{Edge intervalley interaction on graphene half-plane}\label{app:IntervalleyScattering}

Edge intervalley interaction mixes the wave functions from different valleys in BC
\begin{equation}\label{eq:intervalleyBC}
	\left. ( \psi + i g\psi' ) \right|_{at\; edge} = 0.
\end{equation}
Here $\psi$ and $\psi'$ are two component wave functions from the two valleys. The intervalley distance should be explicitly included either in the BC or in the bulk (Weyl-Dirac) equation. We choose the second way and consider modified Hamiltonian, describing a valley,
\begin{equation}\label{eq:intervalleyHam}
	\hat{H} = {\bm\sigma} ({\bm p} - \hbar\bm K_0)
\end{equation}

Then the requirement of absence of normal to the edge current (hermiticity) and the time reversal symmetry \cite{bib:Akhmerov} restricts the matrix $g$ as
\begin{equation}
	g = \frac{ e^{i\phi} }{ \sin\beta} \left( e^{i\gamma \vec\sigma\vec n} + \sigma_3 \cos \beta \right),
\end{equation}
where $\gamma$, $\beta$, $\phi$ are real phenomenological parameters, which can be associated with the parameters from Ref. [\onlinecite{bib:Akhmerov}]. The parameter $\gamma$ is connected with the parameter $a$ as $\sin\gamma = 2a/(1+a^2)$.

Solving the Schrodinger equation with Hamiltonian (\ref{eq:intervalleyHam}) supplemented by the BC (\ref{eq:intervalleyBC}) for the graphene half-plane, we obtain the electronic spectrum, see Fig. \ref{Fig:ESspectra}b. For any $\beta$ and near the projections of the valley centers, $|k_y + sK_{0y}|\ll |K_{0y}|$, the ESs spectrum is expressed as
\begin{equation}
	E_s = s \hbar v\sin\theta (k_y + sK_{0y}), \quad s(k_y + sK_{0y})\cos\theta < 0.
\end{equation}
Here $\cot \theta/2 = (\cos\beta - \cos\gamma)/ \sin\gamma$. Thus, ''rays'' of the ESs start in the $\pm K_{0y}$ points even if intervalley interaction is taken into account. Near the center of the Brillioun zone the rays are transform to each other.

\section{Rippling near edge}\label{app:rippling}

It is well known that ripples in graphene can be treated by means of the effective electrical potential \cite{bib:Kim} $\Phi(\bm{r})=-\kappa_1\left [ (\partial_x^2 h)^2 + (\partial_y^2 h)^2\right ]$ and the vector-potential $\bm{A} = (A_x,A_y)$, where $A_x = -\kappa_2\left [ (\partial_x^2 h)^2 - (\partial_y^2 h)^2\right ] $, $A_y = 2\kappa_2 \left ( \partial_x^2 h + \partial_y^2 h\right )\partial^2_{x,y} h $, $\kappa_1\approx 40 $eV$\cdot$A$^2$, $\kappa_2\approx 6.2$ eV$\cdot$A$^2$. Position of a graphene list in three-dimensional space are determined by the function $h(x,y)$ that determines distance of the list from the plane $z=0$. In this section we study how smooth ripples affect on the ESs spectrum.

First, we consider half-plane $x\geq 0$ when the graphene list is bended in perpendicular to the edge direction (i.e. $\partial_y h=0$). Therefore $k_y$ is a good quantum number. The sfystem of equations for the electron wave function in a valley reads as follows
\begin{equation}\label{Dirac_eq_with_ripple}
	\left\{ 
	\begin{array}{l}
		-\kappa_1\tilde{h}\psi_1+(-i\partial_x - ik_y)\psi_2 - \kappa_2\tilde{h}\psi_2 = E\psi_1 \\
		(-i\partial_x + i p_y)\psi_1 - \kappa_2\tilde{h}\psi_1 - \kappa_1\tilde{h}\psi_2 = E\psi_2
	\end{array}
	\right.,
\end{equation} 
where $\tilde{h}=(\partial_x^2 h)^2$. After multiplying the first equation by $\psi_1$, the second one by $\psi_2$ and subtracting them we get the differential equation for the function $\eta=\psi_1/\psi_2$:
\begin{equation}\label{eta_equation}
	\eta '_{x}-2k_y\eta - i(\kappa_1\tilde{h}+E)(1-\eta^2)=0.
\end{equation}
We solve the equation with the BC (\ref{eq:BC})
\begin{equation}\label{eta_boundary_condition}
	\eta|_{x=0}+ia =0.
\end{equation}
We will perturbatevly treat $\tilde{h}$. Hence we look for the wave function by means of series in powers of $\tilde{h}$, $ \eta = \eta_0 + \eta_1 + \dots $, where $\eta_1 \propto \tilde{h}$ etc. In the zero order $\eta_0 = -i(k_y+\sqrt{k_y^2-E^2})/E$, in the first order we get
\begin{equation}
	\eta_1 = -i\kappa_1(1-\eta_0^2)\int_{x}^{+\infty}\tilde{h}(x')\exp \left\{-2\sqrt{k_y^2-E^2}(x'-x) \right\}dx'.
\end{equation}

Substitution of the first-order wave function in the BC (\ref{eta_boundary_condition}) results in a dispersion equation with the ES spectrum 
\begin{equation}\label{ripple_dispersion}
	\begin{array}{l}
		E = \frac{2a_0k_y}{1+a_0^2}+2\kappa_1k_y\frac{1-a_0^2}{1+a_0^2}\int_{0}^{+\infty}\tilde{h}(x')e^{-2k_y(1-a_0^2) x'}dx',\\
		\\
		\mbox{ïðè} (1-a_0^2)k_y \geq 0.
	\end{array}
\end{equation}
Thus, small edge rippling leads to the renormalization of the parameter $a$. We can obtain spectrum of the ESs in the antidot geometry via quasiclassical quantization of parallel momentum $k_y=l/R$.

\section{Life time of edge states on half-plane with rough boundary}\label{app:roughness}

Let us consider a rough linear edge that can be described by the BC (\ref{eq:BC}) determined on the curve $x=\delta x(y)$, where $\delta x(y)$ is a random deviation of the edge from its average position $\langle \delta x\rangle=0$ in the point $y$, Ref. [\onlinecite{bib:Vasko}]. The edge roughness is assumed to be smooth and $\delta x(y)$ has the Gaussian correlation function $\langle \delta x(y_1) \delta x(y_2)\rangle=\Delta^2 \exp ((y_1-y_2)^2/l_c^2)$. We also assume electron wavelength to be much greater than the correlation length $l_c$.

Firstly, we consider some definite realization of the boundary  $x=\delta x(y)$. Using coordinates transformation $x'=x-\delta x(y), y'=y$ the Hamiltonian (\ref{eq:WeylHam}) with the BC $(\psi_1+ia\psi_2)|_{x=\delta x(y)}=0$ can be transformed to the standart BC $(\psi_1+ia\psi_2)|_{x=0}=0$ (dashes are skipped for brevity) and the same Hamiltonian with additional part
\begin{equation}
\delta \hat H=-v \frac{\partial \delta x(y)}{\partial y} \sigma_y \hat p_{x}.
\end{equation}

The probability of transition from ES with momentum along the edge $\hbar k_y$, wave function $\psi_s$ and energy $E_s$ (for definitness we consider the valey $s=+1$) to bulk state with wave function $\psi_b$ and energy $E_b=\pm v\hbar k'$ due to the edge roughness can be evaluated using the Fermi golden-rule:
\begin{eqnarray}
dw=\frac{2\pi}{\hbar}|\langle \psi_s |\delta \hat H|\psi_b\rangle|^2\delta(E_b-E_s),
\end{eqnarray}

where
\begin{equation}
	\psi_s=\sqrt{\frac{2a^2(1-a^2)}{L_y(1+a^2)^2}k_y}\left( \begin{array}{cc}
        1 \\
        ia^{-1}
    \end{array} \right) e^{-\frac{1-a^2}{1+a^2}k_yx}e^{ik_yy},
\end{equation}

\begin{equation}
	\psi_b=\frac{1}{\sqrt{2L_xL_y}}{}\left( \begin{array}{cc}
        1 \\
        \pm e^{i\phi_{k'}}
    \end{array} \right) e^{ik'_xx}e^{ik'_yy}.
\end{equation}

For simplicity, here we neglect second bulk solution $\sim e^{-ik'_xx}$, because it doesn't sugnificantly affect on final result.

Integrating over ${\bf k}'$ we obtain the probability of ES decay for this realization of the boundary

\begin{multline}
w=\frac{2v}{\pi L_y}\frac{a^2|1-a^2|^3k_y^2}{(1+a^2)^3sgn[a]} \times\\
\int_0^{2\pi}d\phi\left(\frac{1+a^2}{2a}+2sgn[a]\sin \phi\right)\frac{|F(k_y-\frac{2|a|}{1+a^2}k_y\sin \phi)|^2}{(1-a^2)^2+4a^2\cos^2\phi},
\end{multline}
where 
\begin{equation}
F(k)=\int_{-\infty}^{\infty}e^{iky}\frac{\partial \delta x(y)}{\partial y}dy.
\end{equation}

Then we average $w$ over all possible realizations of the boundary using following relation

\begin{equation}
\langle|F(k)|^2\rangle=l_c\Delta^2L_yk^2e^{-l^2k^2/4}.
\end{equation}

Finally, we obtain 
\begin{multline}
\label{Eq:DecayProb}
\langle w \rangle = \frac{2vl_c\Delta^2a^2|1-a^2|^3}{\pi (1+a^2)^3sgn[a]}k_y^2\int_0^{2\pi}d\phi\Biggl[\left(\frac{1+a^2}{2a}+2sgn[a]\sin \phi\right) \\ \frac{(k_y-\frac{2|a|}{1+a^2}k_y \sin \phi)^2}{(1-a^2)^2+4a^2\cos^2\phi} \exp\biggl(-\frac{l_c^2}{4}\left(k_y-\frac{2|a|}{1+a^2}k_y \sin \phi\right)^2\biggr)\Biggr],
\end{multline}

In the limit $a \ll 1$ the probability of decay (\ref{Eq:DecayProb}) has simple form:
\begin{equation}
\langle w \rangle=vl_c\Delta^2|a|k_y^4 e^{-l_c^2k_y^2/4}
\end{equation}

\section{Scattering by charged antidot}\label{app:coulomb}
In the polar coordinates the Weyl-Dirac equation with Coulomb potential can be represented as:
\begin{equation}
	\left( \begin{array}{cc}
        0 & -ie^{-\varphi}\partial_r -\frac{e^{-i\varphi}}{r}\partial_{\varphi} \\
        -ie^{\varphi}\partial_r +\frac{e^{i\varphi}}{r}\partial_{\varphi} & 0 \\
    \end{array} \right)\psi=\left(-k+\frac{q}{r}\right)\psi.
\end{equation}

Following Refs. [\onlinecite{bib:Novikov},\onlinecite{bib:Shytov}], we seek the solution with total angular momentum $j=l+1/2$ for energy $E<0$ in the following form:	
\begin{equation}
\psi^{(l)}(r,\varphi)=
\left( \begin{array}{cc}
\phi(r)+\chi(r) \\
(\phi(r)-\chi(r))e^{i\varphi}
\end{array} \right) r^{s-\frac{1}{2}}e^{il\varphi}e^{ikr},
\end{equation}
where $s=\sqrt{(l+\frac{1}{2})^2+q^2}$.

The functions $\phi(r)$ and $\chi(r)$ are determined by the following expressions:
\begin{equation}
\chi(r)=AM(s+iq,2s+1,-2ikr)+BU(s+iq,2s+1,-2ikr),
\end{equation}
\begin{multline}
\phi(r)=\frac{A(s+iq)}{(l+\frac{1}{2})}M(s+iq+1,2s+1,-2ikr)-\\-B\left(l+\frac{1}{2}\right)U(s+iq+1,2s+1,-2ikr),
\end{multline}
where $M(a,b,z)$ and $U(a,b,z)$ are the Confluent hypergeometric function of the first and second kind respectively. The coefficients $A$ and $B$ are related by the BC:
\begin{widetext}
\begin{equation}
\frac{B}{A}=\frac{(1-ia)\frac{s+iq}{l+\frac{1}{2}}M(s+iq+1,2s+1,-2ikR_0)+(1-ia)M(s+iq,2s+1,-2ikR_0)}{(1+ia)(l+\frac{1}{2})U(s+iq,2s+1,-2ikR_0)-(1-ia)U(s+iq,2s+1,-2ikR_0)}.
\end{equation}
\end{widetext}

To find the renormalization of the quasistationary ES spectrum, we eliminate the incoming wave and obtain the dispersion equation
\begin{multline}
\biggl[(1+ia)\frac{s+iq}{l+1/2}M(s+iq +1,2s+1,z)+\\+(1-ia)M(s+iq ,2s+1,z)\biggr]+\\+\biggl[(1+ia)(l+1/2)U(s+iq +1,2s+1,z)-\\-(1-ia)U(s+iq ,2s+1,z)\biggr]\frac{\Gamma (2s+1)e^{i\pi (-s-iq)}}{\Gamma(s-iq +1)}=0. 
\end{multline}

In the low-energy limit ($kR \ll 1$) and under condition $|q| \ll 1$ for $l\neq 0$ we obtain (\ref{Eq:ESSwithQ}).

Now we turn to the solution of scattering problem. The transport cross section can be expressed in terms of the Coulomb scattering phases $\delta_l$,
\begin{equation}
\sigma_{tr}=\frac{2}{k} \sum _{l=-\infty}^{\infty} \sin^2(\delta_l-\delta_{l+1}),
\end{equation}
where the scattering phases is defined in conventional manner:
\begin{equation}
\label{DefPhas}
\frac{\chi}{\phi}|_{r \rightarrow \infty} = e^{2ikr+2i\beta ln(2kr)-\pi i (l+\frac{1}{2})+2i\delta_l(k)}.
\end{equation}

To find the Coulomb scattering phases we write our solution for $\chi(r)$  and $\phi(r)$ in the limit $r \rightarrow \infty$:

\begin{multline}
\label {Limit1}
\chi(r)\sim A\frac{\Gamma(2s+1)}{\Gamma(s-iq+1)}e^{i\frac{\pi}{2}(s+iq)}(2kr)^{-s}e^{-iq ln(2kr)}+\\+Be^{-i\frac{\pi}{2}(s+iq)}(2kr)^{-s}e^{-iq ln(2kr)},
\end{multline}
\begin{equation}
\label{Limit2}
\phi(r)\sim \frac{A(s+iq)\Gamma(2s+1)}{(l+\frac{1}{2})\Gamma(s+iq+1)}e^{i\frac{\pi}{2}(s-iq)-2ikr+iq ln(2kr)}(2kr)^{-s}.
\end{equation}

Comparing (\ref{DefPhas}) with (\ref{Limit1}) and (\ref{Limit2}) we find
\begin{multline}
e^{2i\delta_l(k)}=\frac{l+\frac{1}{2}}{s+iq}\frac{\Gamma(s+iq+1)}{\Gamma(s-iq+1)}e^{i\pi((l+\frac{1}{2})-s)}+\\+\frac{B}{A}\frac{l+\frac{1}{2}}{s+iq}\frac{\Gamma(s+iq+1)}{\Gamma(2s+1)}e^{i\pi((l+\frac{1}{2})+iq)}
\end{multline}

The results for the transport cross section are presented in Fig. \ref{Fig:CSwithQ}.

\end{document}